\documentclass{article}

\usepackage{arxiv}

\usepackage[utf8]{inputenc}
\usepackage[T1]{fontenc}
\usepackage{hyperref}
\usepackage{url}
\usepackage{booktabs}
\usepackage{amsmath}
\usepackage{amsfonts}
\usepackage{nicefrac}
\usepackage{microtype}
\usepackage{cleveref}
\usepackage{lipsum}
\usepackage{graphicx}
\usepackage{natbib}
\usepackage{doi}

\usepackage{algorithm}
\usepackage{algpseudocode}

\usepackage{listings}
\usepackage{csquotes}
\usepackage{comment}
\usepackage[title]{appendix}

\title{Modeling realistic human behavior using generative agents in a multimodal transport system: Software architecture and Application to Toulouse}

\date{}

\author{\href{https://orcid.org/0009-0000-5354-3998}{\includegraphics[scale=0.06]{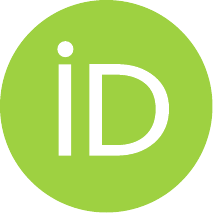}\hspace{1mm}Trung-Dung Vu}\thanks{Corresponding author} \\
	UMR IRIT\\
	University Toulouse Capitole\\
	Toulouse, France \\
	\texttt{vutrungdungbk@gmail.com} \\
	\And
	\href{https://orcid.org/0000-0002-9005-3004}{\includegraphics[scale=0.06]{orcid.pdf}\hspace{1mm}Benoit Gaudou} \\
	UMR IRIT\\
	University Toulouse Capitole\\
	Toulouse, France \\
	\texttt{benoit.gaudou@ut-capitole.fr} \\
	\And
	\href{https://orcid.org/0000-0002-0334-3055}{\includegraphics[scale=0.06]{orcid.pdf}\hspace{1mm}Kamaldeep Singh Oberoi} \\
	CESI\\
	CESI LINEACT\\
	Toulouse, France \\
	\texttt{ksoberoi@cesi.fr} \\
}

\renewcommand{\shorttitle}{Modeling realistic human behavior using generative agents in a multimodal transport system}

\hypersetup{
pdftitle={A template for the arxiv style},
pdfsubject={q-bio.NC, q-bio.QM},
pdfauthor={David S.~Hippocampus, Elias D.~Striatum},
pdfkeywords={First keyword, Second keyword, More},
}

\begin{document}
\maketitle
\begin{abstract}
Modeling realistic human behaviour to understand people's mode choices in order to propose personalised mobility solutions remains challenging. This paper presents an architecture for modeling realistic human mobility behavior in complex multimodal transport systems, demonstrated through a case study in Toulouse, France. We apply Large Language Models (LLMs) within an agent-based simulation to capture decision-making in a real urban setting. The framework integrates the GAMA simulation platform with an LLM-based generative agent, along with General Transit Feed Specification (GTFS) data for public transport, and OpenTripPlanner for multimodal routing. GAMA platform models the interactive transport environment, providing visualization and dynamic agent interactions while eliminating the need to construct the simulation environment from scratch. This design enables a stronger focus on developing generative agents and evaluating their performance in transport decision-making processes. Over a simulated month, results show that agents not only make context-aware transport decisions but also form habits over time. We conclude that combining LLMs with agent-based simulation offers a promising direction for advancing intelligent transportation systems and personalised multimodal mobility solutions. We also discuss some limitations of this approach and outline future work on scaling to larger regions, integrating real-time data, and refining memory models. 
\end{abstract}


\section{Introduction}\label{sec:intro}

Understanding human mobility behavior in multimodal transport systems is essential for advancing traffic planning and the design of efficient transport networks. The challenge lies not only in representing travel patterns but also in building a simulation framework that supports the study of individual behavior, the testing of behavioral hypotheses, and the development of improved multimodal transport and urban traffic planning strategies.

The emergence of personalized multimodal transport systems reflects a broader shift toward user-centric mobility, where public transit, walking, cycling, and shared services are integrated to provide journeys tailored to individual preferences \citep{smith1995transims,horni2016matsim}. Capturing such systems in simulation, however, is particularly challenging because mobility choices are shaped by diverse socioeconomic profiles, accessibility constraints, and personal experiences \citep{grignard2018impact}. As highlighted by \citep{Oberoi2024}, personalization in Mobility-as-a-Service requires addressing the heterogeneity of users—including their preferences, contexts, and accessibility needs—which remains a significant challenge for developing realistic transport models. Modeling, therefore, requires both flexibility and realism to capture heterogeneous and adaptive travel decisions across populations.

Existing transport simulations often struggle with this complexity and with the availability of data needed to calibrate models. In particular, detailed trajectory and itinerary datasets—which are critical for capturing the nuances of multimodal travel behavior—are extremely scarce \citep{fourez2025transport}. Such datasets exist for a limited number of major metropolitan areas, for example, Tokyo, Japan, but are largely unavailable elsewhere \citep{feng2024agentmove}. This dependence on scarce calibration data restricts the applicability of existing models in smaller or less data-rich regions. Moreover, integrating multiple external data sources requires extensive rule engineering, producing brittle models that are difficult to scale and prone to failure in unforeseen conditions.

Recent advances in Large Language Models (LLMs) and generative AI offer a promising alternative \citep{park2023generative_agents}. Generative agents can perceive their environment, process contextual information, learn from experience, and adapt their behavior over time. Beyond transport applications, LLM-based agents have already demonstrated their potential in diverse domains. For example, they have been applied to tasks such as software testing, advertising analysis, and brainstorming in the Tiny Troupe framework \citep{salem2025tinytroupe}, as well as automating research through Agent Laboratory \citep{schmidgall2025agent_laboratory}. These approaches create ensembles of agents with different roles, capabilities, and personas that collaborate to achieve common goals. Additionally, other studies have evaluated the ability of LLM agents to proxy human emotions by testing their performance on psychological tasks grounded in the OCEAN Big Five personality traits \citep{serapio2023ocean}. Results show that these agents can simulate emotional intelligence and empathy, which highlights their potential to proxy human traffic decisions by considering context, preferences, and even inheriting personal traits. At the same time, the GAMA platform provides a powerful environment for spatially explicit agent-based modeling, with strong support for GIS integration and real-world datasets such as GTFS\footnote{\textit{GTFS} is a community-driven open standard for rider-facing transit information, \url{https://gtfs.org/}} and OpenStreetMap\footnote{\textit{OpenStreetMap} is an initiative to create and provide free geographic data, \url{https://osmfoundation.org/}} \citep{taillandier2019gama_platform}.

In this work, we build on these developments with two main objectives: (i) to integrate generative agents with the GAMA\footnote{\textit{GAMA} is an easy-to-use open source modeling and simulation environment for creating spatially explicit agent-based simulations. It has been developed to be used in any application domain: urban mobility, climate change adaptation, epidemiology, disaster evacuation strategy design, urban planning, are some of the application domains in which GAMA users are involved and for which they create models - GAMA Home \url{https://gama-platform.org/wiki/Home}} platform in order to create a scalable, data-driven simulation environment, and (ii) to develop generative agents that act as proxies for human decision-making in multimodal transport systems. Results demonstrate that generative agents can not only adapt their travel choices but also learn and form habits over time, offering a more realistic representation of human mobility behavior. Nevertheless, the well-known limitations of LLMs—such as inconsistency, hallucination, grounding in real-world data, and scalability—remain significant challenges \citep{taillandier2025llm_abm}. Addressing these issues constitutes an important direction for our future research.

The remainder of this paper is organized as follows: Section \ref{sec:related} reviews related work on the integration of LLM-based generative agents into urban mobility simulation. Section \ref{sec:methodology} presents the proposed methodology, including the software architecture, generative agent components, and GAMA platform integration. Section \ref{sec:experiments} describes the experimental setup and presents the results. Finally, Sections \ref{sec:discussion} and \ref{sec:conclusion} discuss the contributions, limitations, and directions for future research.

The code is available at \url{https://github.com/dungzvu/llm-agents-gama}.

\section{Related Work}\label{sec:related}

\subsection{The Conceptual Framework}\label{sub:related:framework}

The notion of generative agents has gained traction from the influential work of \citep{park2023generative_agents}, which describes a core architecture for artificial agents endowed with memory, planning, observation, and reflection modules. In this architecture, agents record experiences in a memory stream, manage and summarize memories through reflection, and reference these memories to guide future behavior via planning and reasoning. These core modules—perception, memory (both short-term and long-term), planning, and reflection—form the backbone of autonomous, context-aware agent behavior. In this work, generative agents leverage LLMs to generate behavior and interact with one another in a human-like manner. Building on this foundation, subsequent studies have extended the generative agent paradigm to various domains. For instance, \textit{Voyager} is an agent that continuously explores the world, learns, and makes novel discoveries without human intervention, demonstrated within the Minecraft environment \citep{wang2023voyager}. Similarly, \textit{TinyTroupe} and \textit{XAgent} enable the simulation of groups of people collaborating to achieve shared goals \citep{salem2025tinytroupe, xagent2023}.

Extending this architecture into the domain of human mobility, \citep{liu2024llm_agent_based_modeling} propose a conceptual framework tailored to transportation systems, grounded in the generative agent paradigm. Their framework augments traditional modules with transport-specific components, including general planning, travel planning, and dual memory layers (long-term and short-term memory). It also incorporates personal identity and traits as inputs, allowing agents' decisions to be influenced by their unique socio-demographic and behavioral profiles. In this setup, LLMs are prompted with identity- and trait-based inputs, enabling them to act as proxies for human decision-makers. The models generate textual outputs that represent decisions, which are then executed by agents in the simulation environment. In their demonstration on a simplified transport network divided into four zones, agents use multi-layer travel planning and experiential learning to iteratively optimize travel routes, which helps alleviate traffic congestion in that small network.

This framework effectively adapts the generative agent architecture to mobility contexts, enabling agents to reason over high-level intentions (general plan) and concrete route choices (travel plan), while learning from experience and adjusting behavior over time. Provides a conceptual baseline for integrating generative agents into transport modeling by embedding agent cognition (memory, planning, trait-driven decisions) within a networked mobility environment.

\subsection{City-scale Mobility Simulation}\label{sub:related:simulation}

One promising direction in mobility modeling is using generative or agentic frameworks to predict or simulate city-scale movement. A recent effort, called AgentMove \citep{feng2024agentmove} combines LLM reasoning with spatial modeling to forecast agents' next destinations. This work casts the city as a multi-layer graph to encode spatial context and decomposes the prediction task into modules for spatio-temporal memory, world knowledge generation, and collective pattern extraction. While this approach achieves strong predictive accuracy on benchmark datasets, it relies on rich trajectory and survey data that are rarely available in most cities, which limits its general applicability.

A more traditional but practical approach is demonstrated by Ile-de-France's synthetic itinerary generation (via the EQASIM framework) \citep{horl2021synthetic}. This method uses population data, origin-destination matrices, and statistical matching techniques to generate daily itineraries for synthetic agents. Although less expressive in modeling individual cognition, it is grounded in widely available demographic and commuter data and can be applied even in regions lacking fine-grained trajectory logs.

At a larger scale, the \textit{AgentSociety} framework is designed to simulate 10,000 agents in an urban environment \citep{piao2025agentsociety}. This work primarily focuses on optimizing the underlying architecture to support large-scale populations, enabling agents to form daily routines, interact, and influence one another in socially meaningful ways. However, it does not go into the complex of transport systems.

Another recent contribution is GATSim\citep{liu2025gatsim} which introduces a generative-agent architecture over a simplified transport network consisting of 16 fully connected nodes in a single-modal setting. Agents in GATSim are equipped with memory hierarchies, planning and reflection mechanisms, and learning capabilities, which allow them to generate realistic travel behaviors. The framework demonstrates that emergent traffic patterns align with expected macroscopic dynamics (such as peak spreading), and that agent trajectories reproduce aspects of human-like decision-making. However, its reliance on a small, simplified network and single-mode transport limits its representativeness for complex urban environments.

Taken together, these works map out a design space between pure prediction (AgentMove), synthetic generation (Ile-de-France), and full simulation (GATSim, AgentSociety). They illustrate both the potential and the barriers of applying the human traveling behavior simulation at city scale. In the next section, we will position our approach in this space, aiming to bridge toward multimodal, large-scale transport simulation with generative agents.

\begin{figure}
	\centering
	\includegraphics{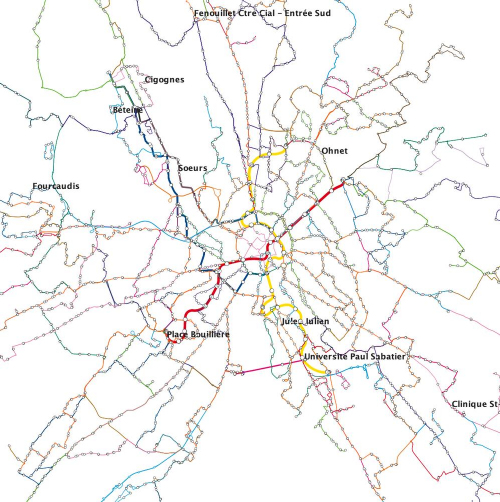}
	\caption{The public transport system of Toulouse, France, combining metro, tram, bus, and cable car, visualized in the GAMA platform with GTFS data.}
	\label{fig:toulouse_transport_system}
\end{figure}

\subsection{Research Gap and Opportunities}\label{sub:related:gap}

Despite recent advances, current generative-agent approaches remain limited to simplified or small-scale transport networks, restricting their relevance for real multimodal systems. To be practically useful, agents must operate in complex environments where multiple modes and dense urban structures interact. For example, the public transport system of Toulouse— combining metro, tram, bus, and cable car — illustrates such a setting, where realistic decision-making requires agents that adapt travel choices to context and experience (Figure \ref{fig:toulouse_transport_system}); the transport network is far more complex than the 16 fully connected nodes used in GATSim, as it features a dense structure with numerous stops and routes, integrates multiple modes, and can be represented as a large multi-layer graph, where each stop and route corresponds to a node and an edge, respectively.

Another gap lies in the construction of simulation environments. Many studies develop platforms from scratch, which demands significant coding effort and hinders scalability. Integrating generative agents with the established GAMA platform provides a more efficient path, allowing researchers to focus on agent performance while leveraging GAMA’s built-in spatial modeling, visualization, and data integration capabilities. The GAMA platform is particularly well-suited for multi-agent, spatially explicit transport simulations, as it supports seamless import of GIS datasets such as shapefiles and OpenStreetMap (OSM) data, and can simply agentify spatial features (e.g., roads, buildings) to represent transport networks and urban environments in detail. Other than that, no prior work has explored the integration of LLM-based generative agents with the GAMA platform at the time of this study.

Addressing these gaps creates opportunities to advance city-scale simulations by embedding generative agents in multimodal transport networks, enabling richer behavioral insights and supporting sustainable urban mobility planning.

\section{Methodology}\label{sec:methodology}

\subsection{Overall Architecture}\label{sec:methodology:overall_architecture}

\begin{figure}
  \centering
  \includegraphics{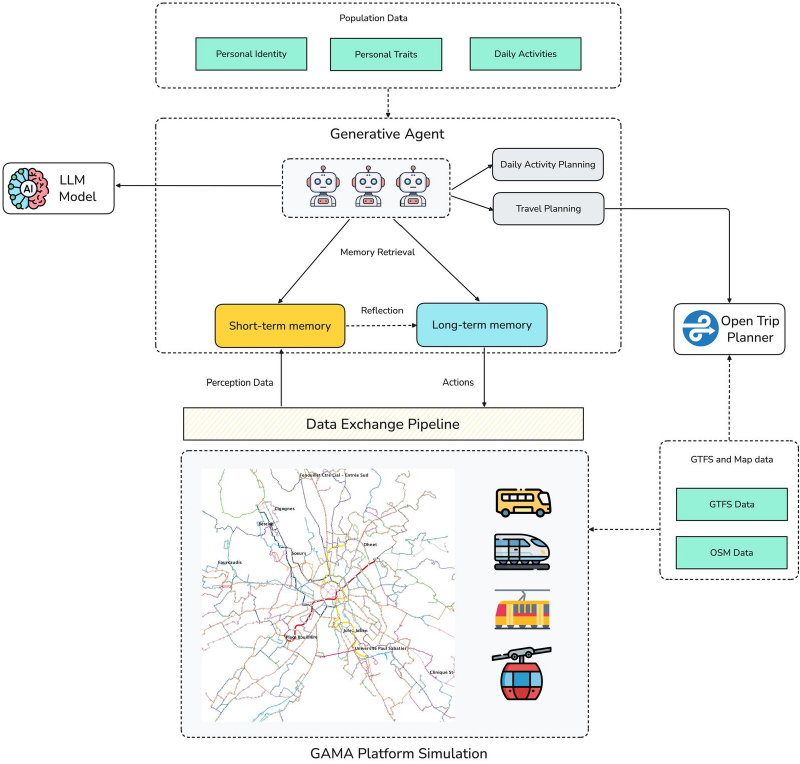}
  \caption{The architecture of the proposed framework, integrating the GAMA platform with Generative Agents and the OpenTripPlanner module.}
  \label{fig:method_architecture_overview}
\end{figure}

The proposed architecture builds upon the conceptual framework introduced by \citep{liu2024llm_agent_based_modeling}. Its core components are: (i) a multimodal transport system modeled in the GAMA Platform, (ii) Generative Agents, and (iii) a Data Exchange Integration layer that connects these two modules. Through this integration layer, the Generative Agents control and coordinate the behavior of agents within the GAMA simulation, while continuously adapting their future decisions according to evolving contexts and feedback. An overview of the architecture is illustrated in Figure \ref{fig:method_architecture_overview}.

The architecture relies on a diverse set of input data. At the individual level, each agent is initialized with a \textit{personal identity and traits} as well as a set of \textit{daily activities} represented as a sequence of actions that specify the time, location, and purpose (e.g., attending school, commuting to work, engaging in leisure, or returning home). At the system level, the architecture integrates \textit{geographic data} and \textit{GTFS data}, the latter providing the schedules of the city’s public transport system.

Decision-making is supported by the \textit{Generative Agents}, which query LLM models to generate context-aware plans and adapt them to real-time conditions. A dedicated \textit{Daily Activity Planning} module (i) translates each agent’s activity sequence into a structured schedule, (ii) triggers activities at their designated times, and (iii) adjusts departure times to ensure timely arrivals. For example, if an agent departs for work at 8:00 AM on the first day but arrives 20 minutes late, it is likely to adjust its departure time to 7:40 AM on the following day. The \textit{Daily Activity Planning} module then invokes the \textit{Travel Planning} module, which decomposes each activity transition into a detailed journey, specifying travel legs, selecting mobility modes, and determining the routes to be followed. Within this setup, perception data—including travel outcomes such as delays, waiting times, in-vehicle travel times, and arrival deviations—are transmitted to the Generative Agent memory in real time. This enables agents to update their short-term observations and long-term reflections as the simulation progresses.

In this architecture, each agent in the GAMA platform simulation corresponds to a generative agent. We extend the original framework of \citep{liu2024llm_agent_based_modeling} by incorporating an additional OpenTripPlanner module. The module provides access to detailed transit data and multimodal routing information, enabling Generative Agents to evaluate alternative journeys and adapt their travel plans dynamically. More importantly, it allows agents to operate within the complex topology of transport networks, represented as large multi-layer graphs with numerous stops, routes, and modes of transport. By embedding this capability into the architecture, agents are able to replicate human decision-making more realistically: consulting route information, responding to disruptions, and selecting journeys that best align with their preferences and context.

To summarize, \textit{Generative Agents}, supported by various modules and data sources, leverage LLMs to emulate human behavior based on personal identity and traits. With the integration of \textit{OpenTripPlanner} as a travel assistant, agents can navigate the simulation environment and complete their daily activities without being constrained by the inherent complexity of transport networks. Finally, the GAMA platform provides the simulation environment in which agents learn, interact, and act across diverse experimental scenarios.

\subsection{Simulation in GAMA}\label{sec:methodology:simulation_in_gama}

In the simulation environment, the transport network is constructed by integrating GTFS data, which includes information about routes, stops, schedules, and other relevant details. The simulation platform GAMA natively supports modeling and simulating spatial environments, making it easy to visualize the transport network and the movement of agents within it. We define two main types of agents: \textit{Inhabitant} agents and \textit{Public Vehicle} agents, which respectively represent the individuals making travel decisions and the vehicles operating within the simulated transport system.

While \textit{Public Vehicle} agents follow predefined routes and schedules based on the GTFS data, \textit{Inhabitant} agents are controlled by their corresponding Generative Agents, and also send the perception data back to the Generative Agents in real time. Algorithm \ref{alg:inhabitant_behavior} shows the behavior of the \textit{Inhabitant} agent running in GAMA platform, given a list of actions (i.e., legs) generated by its corresponding Generative Agent. Each leg represents either a \textit{transfer}—movement between stations or bus stops—or a \textit{transit}—travel using an available public transport vehicle. The agent iterates through the list of actions, executing each one in sequence. Upon arriving at the destination stop, the agent disembarks and proceeds to the next action. The process continues until all actions have been completed, at which point the agent submits a final perception data indicating arrival and resets its action list.

\begin{algorithm}
    \caption{Inhabitant Behavior}
    \label{alg:inhabitant_behavior}
    \begin{algorithmic}[1]
        \Require{$ ACTIONS = leg_1, leg_2, \ldots, leg_N$}
        \Statex

        \State $index \gets 0$

        \State \textbf{GAMA Main Loop}
        \While{$SIMULATION\_RUNNING$}
            \If{IsTransfer(ACTIONS[index])}
                \State MoveToStop(ACTIONS[index].to\_stop)
                \State $index \gets index + 1$
            \Else
                \State $Closest \gets ClosestVehicle(ACTIONS[index].mode)$
                \If{CanGetIn(Closest)}
                    \State \textit{SubmitPerceptionData}(WAITED\_FOR\_VEHICLE)
                    \State GetInVehicle(Closest)
                \EndIf
            \EndIf

            \Statex
            \If{ArrivedAtDestination(ACTIONS[index].to\_stop)}
                \State \textit{SubmitPerceptionData}(TRANSIT\_DURATION)
                \State $GetOffVehicle()$
                \State $index \gets index + 1$
            \EndIf

            \Statex
            \If{$index \geq N$}
                \State \textit{SubmitPerceptionData}(ARRIVED)
                \State $ACTIONS \gets NULL$
            \EndIf
        \EndWhile
    \end{algorithmic}
\end{algorithm}

Algorithm \ref{alg:llm_gas_server_main_loop} shows the main simulation loop, which is running in Generative Agent component. The main loop consists of three key components: (i) synchronously waiting for GAMA to send the /sync request, and timely send \textit{actions} to GAMA - which are the input of the Algorithm~\ref{alg:inhabitant_behavior}, (ii) performing memory reflection at the end of each day, and (iii) processing incoming perception data from GAMA asynchronously. The synchronous waiting ensures that the Generative Agents have the most up-to-date state of the agents in GAMA before making any decisions. The memory reflection allows agents to consolidate daily experiences into structured knowledge that can be reused in subsequent decisions. The asynchronous processing of perception data enables the Generative Agents to receive updates on the outcomes of their actions in real time, which is crucial for learning and improving decision-making over time.

\begin{algorithm}
    \caption{Main Loop}
    \label{alg:llm_gas_server_main_loop}
    \begin{algorithmic}[1]
        \State \textbf{Main Loop}
        \While{$SIMULATION\_RUNNING$}
            \State \textbf{Synchronously wait for GAMA to send the /sync request}
            \State $idlePeople, timestamp \gets GetSimulationStateFromGAMA()$
            \State $actions \gets GetNextActionsForInhabitants(idlePeople, timestamp)$
            \State $SendActionsToGAMA(actions)$
            \Statex
            \State \textbf{Do memory reflection at the end of each day}
            \If{EndOfDay(timestamp)}
                \For{$agent, memories \in GetAllShortTermMemories()$}
                    \State $ReflectAgentMemory(agent, memories)$
                \EndFor
            \EndIf

            \Statex
            \State \textbf{Feedback Loop, process incoming perception data from GAMA asynchronously}
            \For{$perceptionData \in ReceivePerceptionDataFromGAMA()$}
                \State $agent \gets GetAgent(perceptionData.personID)$
                \State $AppendShortTermMemory(agent, perceptionData)$
            \EndFor
        \EndWhile
    \end{algorithmic}
\end{algorithm}

To be more specific, in the GAMA platform we simulate the public transport system of Toulouse, which consists of nearly 450 concurrent vehicles (buses, metro, tram, and cable car) and 5 inhabitant agents. These agents represent employed workers whose identities and daily activities are generated from \textit{Île-de-France} data and are randomly assigned diverse personal traits. Further details of this setup are provided in Section \ref{sec:experiments:setup}.

\subsection{Tackling Network Complexity}\label{sec:methodology:tackle_network_complexity}

In this work, we aim to simulate human mobility behavior in a complex multimodal transport system, which can be consider as a large multi-layer graph, where each stop and route corresponds to a node and an edge, respectively. The complexity arises from the dense structure with numerous stops and routes, the integration of multiple modes of transport (e.g., metro, tram, bus, cable car), and the dynamic nature of real-world transit systems (e.g., schedules, delays, disruptions). 

To handle the complexity of the transport network, we leverage the OpenTripPlanner\footnote{\textit{OpenTripPlanner (OTP)} is an open-source software that provides routing services using GTFS and OpenStreetMap data, \url{https://www.opentripplanner.org/}} module to provide multimodal routing capabilities. OpenTripPlanner is an open-source platform that supports various modes of transport, including walking, cycling, driving, and public transit. It can process General Transit Feed Specification (GTFS) data to generate detailed route options based on real-world transit schedules and routes.

\begin{figure}
  \centering
  \includegraphics{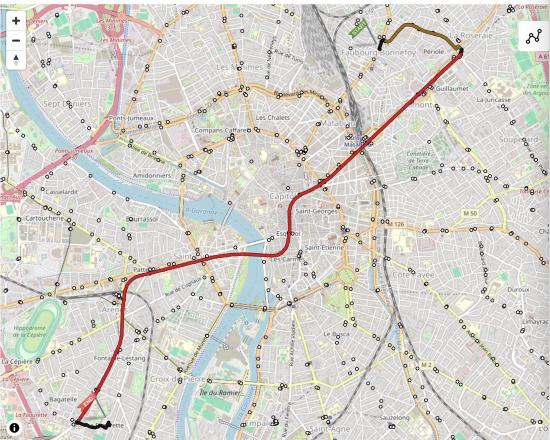}
  \caption{An example of multimodal routing using OpenTripPlanner. The map shows the public transport network in Toulouse, France.}
  \label{fig:OpenTripPlanner_example}
\end{figure}

Figure \ref{fig:OpenTripPlanner_example} shows an example of multimodal routing using OpenTripPlanner. The map displays the public transport network in Toulouse, France, which includes metro, tram, bus, and cable car lines. The output of OpenTripPlanner includes multiple route options, each with detailed information such as departure and arrival times, transfer points, the estimated walking distances, and travel durations. This information then is translated into a textual format that can be understood by Large Language Models as shown below:

\begin{lstlisting}[caption={Example of OpenTripPlanner itinerary in textual format}, label=list:OpenTripPlanner_itinerary_text_example, numbers=none, frame=single]
List of transits:
- Bus 36 from 'Hauts de Bonnefoy' to 'Roseraie'; estimated time: 4 minutes
- Metro A from 'Roseraie' to 'Bagatelle'; estimated time: 15 minutes
Expected starting in 3 minutes.
Total walking time between transits: 15 minutes; 
Total walking distance: 757.0 meters
\end{lstlisting}

The role of \textit{OpenTripPlanner} in constructing the prompt will be further discussed in Section \ref{sec:methodology:evolving_learning_reflection}.

\subsection{Generative Agent Memory}\label{sec:methodology:agent_memory}

The memory system is central to the agent's ability to learn from experience and adapt behavior over time. This module takes a crucial role in storing, retrieving, and managing knowledge at both the naive and the reflective levels, joining both reflection and decision-making phases. How the agent performs depends heavily on the design of this memory system. Like other related works, we implement a dual-layer memory architecture consisting of short-term and long-term memory components.

\subsubsection{Short-term Memory}\label{sec:methodology:short_term_memory}

The short-term memory is designed to store recent experiences and observations in a relatively simple form. It captures the agent's immediate context and the events occurring in its surroundings. For example, in the current implementation, the short-term memory includes details of recent trips, such as the \textit{delays, waiting times, travel durations}, and \textit{arrival late time}. In future work, this memory could be extended to include additional contextual factors—such as weather conditions, route disruptions, or traffic incidents—to support experimentation under varied scenarios. Short-term memory is volatile and primarily serves as input to the reflection process at the end of each simulated day. Structurally, it is organized as a chronological stream of events, allowing agents to review and reason over daily experiences in sequence.

\subsubsection{Long-term Memory}\label{sec:methodology:long_term_memory}

The long-term memory is responsible for storing knowledge consolidated from the short-term memory over time. This includes successful strategies, learned preferences, and other insights that inform future decision-making. Unlike short-term memory, it is less volatile and designed to retain information across multiple simulation runs, enabling the agent to develop a richer understanding of its environment and progressively improve performance.

To maximize the utility of information derived from short-term memory, we define two distinct types of long-term memory: (i) \textit{concepts}, which represent specific events or facts independent of personal traits or identity, and (ii) \textit{reflections}, which summarize broader insights and lessons learned, often incorporating the agent's traits and identity. Concepts serve as concise representations of short-term memory, optimizing token usage in large language model prompts, while reflections capture higher-level abstractions and patterns that can guide future behavior in alignment with the agent's characteristics.

For example, a concept might be \textit{\enquote{Bus 67 tends to be slow during the evening commute}}. A reflection, however, could vary depending on the agent's traits: one individual might prefer this bus because of its predictability, while another with high conscientiousness might avoid it due to its slowness, even though both agents have learned the same underlying concept.

Another type of long-term memory involves abstracting the \textit{reflections} into higher-level generalizations, which are discussed further in Section~\ref{sec:experiments}.

\subsubsection{Memory Retrieval}\label{sec:memory_retrieval}

\begin{figure}
\centering
\includegraphics{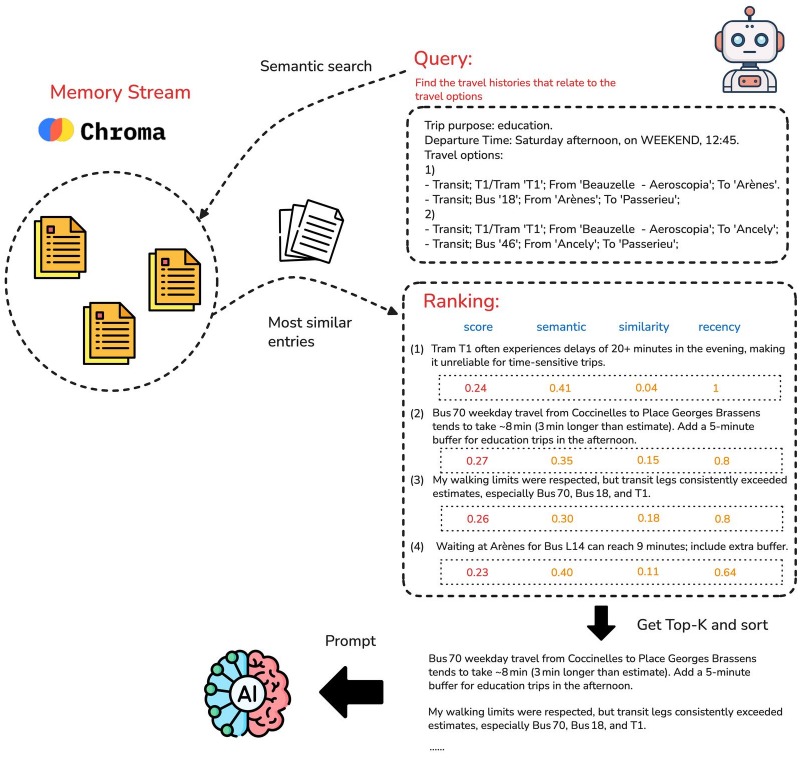}
  \caption{The memory stream consists of historical experiences that may be relevant to the current transport context. A retrieval step identifies a subset of these experiences to be provided to the language model, enabling more informed selection among the available travel options.}
  \label{fig:method_memory_retrieval}
\end{figure}

The memory retrieval process (Figure \ref{fig:method_memory_retrieval}) is crucial for the agent's decision-making capabilities. To make it realistic, we aim to mimic the human cognitive process of recalling relevant information from memory when faced with a decision. Our approach integrates:
\begin{itemize}
    \item \textit{Time decay}: Recent memories are more accessible than older ones, reflecting the human tendency to prioritize recent experiences.
    \item \textit{Relevance}: Memories that are contextually relevant to the current situation are prioritized, ensuring that the agent retrieves information that is most likely to inform its decisions effectively. We consider two kinds of \textit{Relavance}:
    \begin{itemize}
      \item \textit{Contextual relevance}: Memories related to the current context (e.g., location, time of day) are prioritized.
      \item \textit{Seasonal relevance}: Because the transport system can be affected by seasonal changes (e.g., holidays, rush hours, weekdays and weekends), memories from similar periods in the past are given higher priority.
    \end{itemize}
\end{itemize}

For \textit{contextual relevance} retrieval, we use Chroma\footnote{\textit{Chroma} is the open-source search and retrieval database for AI applications, \url{https://www.trychroma.com/}} - a vector database to store long-term memories as high-dimensional vectors. Each memory entry is encoded using a pretrained language model (the \textit{all-MiniLM-L6-v2} model from Sentence Transformers) to capture semantic meaning. When the agent needs to retrieve memories, it generates a query vector based on the current context and searches the vector database for the most similar entries. After that, the retrieved memories are ranked based on \textit{time decay} and \textit{seasonal relevance} criteria to ensure that the most pertinent information is prioritized.

To be more specific, the ranking score for each memory entry is calculated as follows. First, we calculate the $semantic\_score$ by measuring the cosine similarity between the query vector and each memory vector. The query represents a short sequence describing the context of the upcoming activity. For example, this context may include the activity’s time, location, and purpose, as well as the sequence of stops and routes suggested by OpenTripPlanner.
\begin{equation}
    semantic\_score(\mathbf{c}, \mathbf{r}) = \frac{\mathbf{c} \cdot \mathbf{r}}
                                                   {\|\mathbf{c}\| \, \|\mathbf{r}\|}
\end{equation}
where $\mathbf{c}$ and $\mathbf{r}$ are the embedding vectors of the memory entry and the query, respectively.

Next, the $similarity\_score$ is calculated based on the \textbf{BLEU-2} score as
\begin{equation}
    similarity\_score = \textbf{BLEU-2}(C, R)
\end{equation}
where $C$ is the content of the memory entry and $R$ is the query. We use this score to heuristically deal with the \textit{seasonal relevance} criterion, while also boosting the \textit{contextual relevance} criterion. For instance, the seasonal relevance can be captured by keywords such as \enquote{weekend}, \enquote{holiday}, or \enquote{rush hour}, the time range like \enquote{10 AM - 2 PM}, or specific day of week, and the keywords like \enquote{weekdays} or \enquote{weekends}.
    
We calculate also the $recency\_score$ based on the time stamp of the memory, using an exponential decay function as
\begin{equation}
    recency\_score(t) = \lambda^{(T - t)}     
\end{equation}
where $T - t$ is the number of days since the memory was created and $\lambda$ is a decay rate parameter that controls how quickly the score decreases over time.

Finally, the $overall\_score$ is calculated as a weighted sum of $semantic\_score$, $similarity\_score$, and $recency\_score$:
\begin{equation}
    overall\_score = \alpha \cdot norm(semantic\_score) + \beta \cdot norm(similarity\_score) + \gamma \cdot norm(recency\_score)
\end{equation}
where $\alpha$, $\beta$, and $\gamma$ are the weights assigned to each score component respectively, and $norm()$ is a normalization function that ensures the scores are on a comparable scale. This $overall\_score$ is used to rank the memories and select the top-$k$ most relevant ones for the current query.

Figure \ref{fig:method_memory_retrieval} illustrates the memory retrieval process. First, the next activity is triggered by the \textit{Daily Activity Planning} module, which specifies the travel purpose, planned departure time, and source and destination locations. Next, travel options are added to the context by querying OpenTripPlanner. The generative agent then performs a semantic search to retrieve the most similar memory entries from the Memory Stream database (Chroma in this implementation). These entries are ranked using the scoring system described above, and the top-k entries are selected and incorporated into the prompt that is sent to the LLM to support decision-making.

Looking more closely at the memory entries in Figure \ref{fig:method_memory_retrieval}, the 1st and 4th entries achieve the highest $semantic\_score$ since they relate to experiences with \textit{Tram T1} and the \textit{Arènes} bus stop. However, the 1st entry does not yield a high $similarity\_score$ because it occurred in the \textit{evening}, whereas the current context is in the \textit{afternoon}. The 2nd entry, by contrast, has a lower $semantic\_score$ but a higher $similarity\_score$, as it corresponds to an \textit{afternoon} route for \textit{education}, which aligns more closely with the context. This demonstrates how $semantic\_score$ and $similarity\_score$ complement each other in identifying the most relevant experiences, while the $recency\_score$ ensures that more recent experiences are weighted more heavily than older ones.

\subsection{GAMA Platform Integration}\label{sec:methodology:gama_integration}

\begin{figure}
\centering
 \includegraphics[scale=1.1]{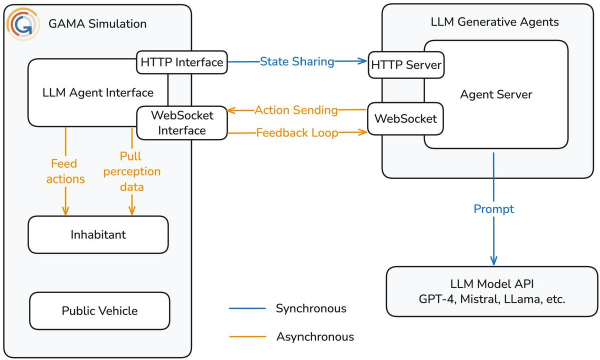}
  \caption{Data exchange pipeline between GAMA and LLM GAs.}
  \label{fig:data_exchange_pipeline}
\end{figure}

The integration of LLM–based Generative Agents with the GAMA simulation environment is achieved through a combination of synchronous and asynchronous communication mechanisms, specifically HTTP and WebSocket protocols. The details of this integration are shown in Figure \ref{fig:data_exchange_pipeline}. Since GAMA platform has built-in support for these protocols, data can be exchanged between the two systems in real time. Within GAMA, a dedicated model, referred to as the \textit{LLM Agent Interface}, serves as the communication bridge to the Generative Agent Server, which is implemented as a Python-based HTTP and WebSocket server.

Building on this integration layer, the Generative Agent (GA) itself is implemented outside of the GAMA simulation platform in order to take advantage of the rich Python ecosystem. While GAMA provides strong capabilities for modeling and simulating spatial environments, its native language (GAML) does not offer support for essential libraries commonly used in large language model development, such as retrieval-augmented generation systems, vector databases, and advanced machine learning frameworks.

By hosting the Generative Agent as a separate Python-based server application, we encapsulate all its functionalities and expose them through a set of application programming interfaces for communication with the GAMA environment. This design ensures that the simulation benefits from GAMA's strengths in spatial modeling while still enabling advanced reasoning, memory management, and decision-making capabilities powered by the Python ecosystem. The integration process involves several key steps:
\begin{itemize}
    \item \textbf{Action Sending:} The first step is to send the actions - the sequence of \textit{transfer} and \textit{transit} actions (see Section~\ref{sec:methodology:simulation_in_gama}) defined in the LLM Generative Agents to the GAMA simulation. Whenever an inhabitant agent in GAMA receives a new action command from its corresponding LLM GA, it executes the action within the simulation environment.
    
    \item \textbf{State Sharing:} Because the agents in GAMA have a one-to-one mapping with their corresponding agents in the LLM GAs, the state of each agent in GAMA is shared with its counterpart in the GAs. This allows the GAs to have a consistent view of the simulation state and make informed decisions based on the current context. The shared state includes (i) the \textit{simulation timestamp} and (ii) the \textit{spatial location} of each agent. A critical aspect of this process is timestamp synchronization: because it takes time for the GAs to process the information and respond to the simulation, so we need to delay GAMA to force it to wait.
    
    \item \textbf{Feedback Loop:} A feedback loop is established between the GAs and the GAMA simulation, allowing the GAs to continuously receive updates on the outcomes of their actions. This feedback is crucial for the GAs to learn and improve their decision-making over time. Specifically, the feedback corresponds to the observations described in Section~\ref{sec:methodology:short_term_memory}.
\end{itemize}

\textit{State Sharing} step is performed in a synchronous manner, because the GAs need to have the most up-to-date state of the agents in GAMA before making any decisions. On the other hand, the \textit{Action Sending} and \textit{Feedback Loop} steps are handled asynchronously, as the GAs can process actions and feedback independently of the simulation's main loop. Therefore, \textit{State Sharing} step is implemented using HTTP protocol, while \textit{Action Sending} and \textit{Feedback Loop} are implemented using WebSocket protocol to improve the speed of the simulation.

\subsection{Evolving Learning and Reflection}\label{sec:methodology:evolving_learning_reflection}

\begin{figure}
\centering
  \includegraphics[scale=0.96]{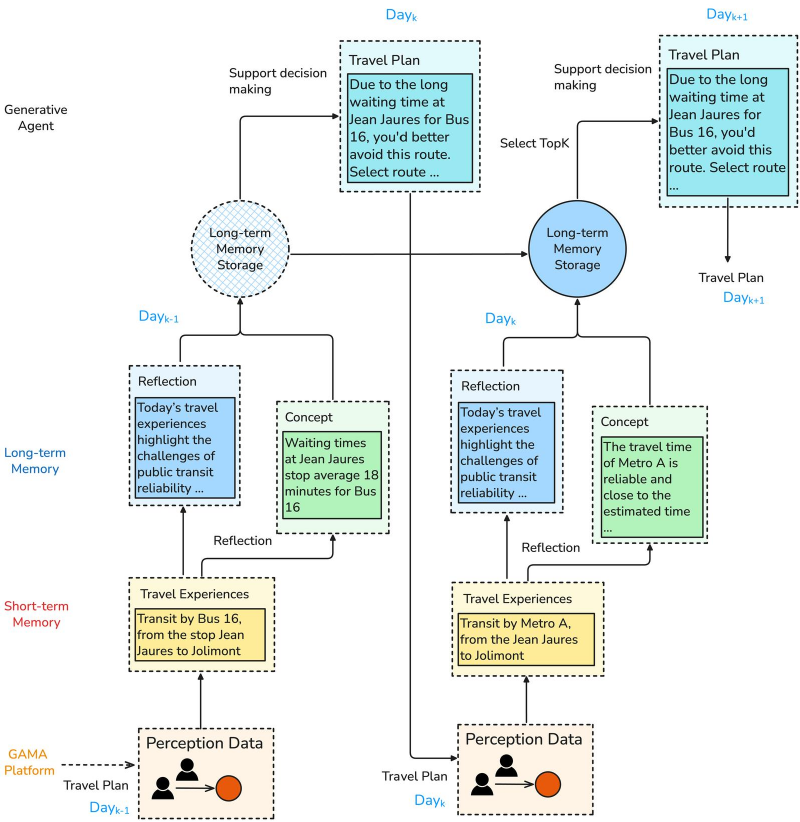}
  \caption{The reasoning flow of the Generative Agent.}
  \label{fig:reasoning_flow}
\end{figure}

The reasoning flow of the Generative Agent, illustrated in Figure \ref{fig:reasoning_flow}, operates through two interconnected phases that enable agents to adapt their travel behavior over time:

\begin{itemize}
  \item \textbf{Reflection phase} — At the end of each simulated day, the agent reviews perception data (such as delays, waiting times, and travel outcomes) stored in short-term memory and integrates this information into long-term memory. This process consolidates daily experiences into structured knowledge that can be reused in subsequent decisions.
  
  \item \textbf{Decision-making phase} — At the beginning of the next simulated day, the agent retrieves relevant knowledge from long-term memory to inform travel planning. The plan is shaped by daily activities, personal identity, individual traits, and accumulated experience. To select among alternatives, the agent queries the OpenTripPlanner module to obtain multimodal route options across the transport network; these candidate itineraries are then evaluated against the agent's preferences and constraints. The finalized itinerary is then transmitted to the corresponding inhabitant agent in the GAMA simulation environment for execution.
\end{itemize}

This iterative cycle allows agents to refine their choices across simulation days. For example, as depicted in Figure \ref{fig:reasoning_flow}, an agent may learn on day $D_{k-1}$ that waiting times at the Jean Jaurès stop average 18 minutes for Bus 16. On day $D_k$, the agent applies this knowledge by avoiding Bus 16 and selecting an alternative route. The mechanism is conceptually similar to a Recurrent Neural Network (RNN) in natural language processing, where outputs from the step $D_{k-1}$ serve as inputs to the next step $D_k$. Here, however, continuity is maintained by the explicit memory management system, and the objective is to generate adaptive travel plans rather than text sequences.

\begin{figure}
\centering
  \includegraphics{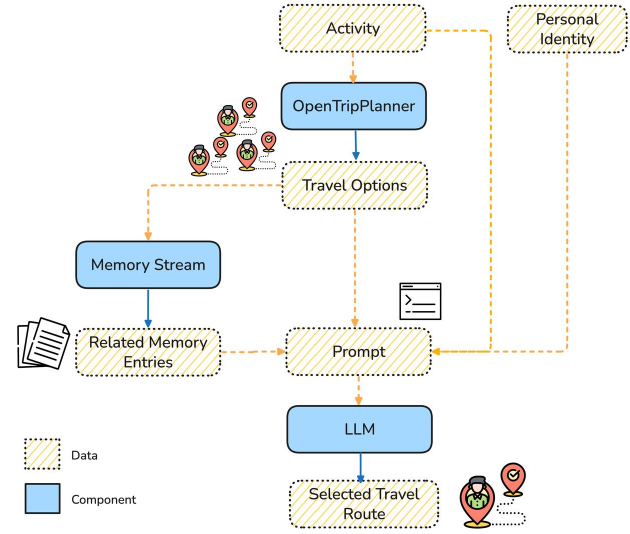}
  \caption{The data flow for constructing a prompt to be used with the LLM.}
  \label{fig:prompt_builder}
\end{figure}

Figure \ref{fig:prompt_builder} illustrates the data flow for constructing a prompt used with the LLM—the core engine of the Generative Agent. The process begins with an \textit{activity} that is triggered at a specific time. Next, \textit{OpenTripPlanner} is queried to provide several candidate \textit{travel options}. The \textit{activity} and \textit{travel options} are then combined and sent to the \textit{memory stream} (long-term memory) to retrieve \textit{related memory entries}, as discussed in Section \ref{sec:memory_retrieval}. Finally, four elements are integrated to build the prompt: (i) the \textit{activity}, (ii) the agent’s \textit{personal identity}, (iii) the \textit{related memory entries}, and (iv) the candidate \textit{travel options}. This prompt is then passed to the LLM, which selects the most suitable travel route for the agent in the GAMA simulation.

\section{Experiments}\label{sec:experiments}

\subsection{Experimental Setup}\label{sec:experiments:setup}

The experiments were conducted using the public transport system of Toulouse, France. The simulation environment was constructed from General Transit Feed Specification (GTFS) data published in March 2025 and complemented with geographic layers derived from OpenStreetMap (OSM). The month of March was chosen since there were no public holidays in France during this month which made it easier to model cyclic and repetitive behaviour of agents. The study simulated a population of 5 generative agents. Each agent was assigned synthetic data describing daily activity patterns and personal traits. Daily activities were adapted from the publicly available \textit{Île-de-France} mobility dataset for Toulouse\footnote{A guide to generate the synthetic data for Toulouse, France \url{https://github.com/eqasim-org/ile-de-france/blob/develop/docs/cases/toulouse.md}}, France \citep{horl2021synthetic}. Personal traits were generated with ChatGPT (\textit{GPT-4o} model) following the structured persona format introduced in the \textit{TinyTroupe} project \citep{salem2025tinytroupe}. 
The detailed data are presented in \ref{tab:population_data}.

For the generative component, we used several open-source large language models available on the Groq\footnote{\textit{Groq} is a platform for hosting large language models (LLMs), providing high-speed and cost-efficient inference, \url{https://groq.com/}} platform: \textbf{GPT OSS 120B (128k context window)}, \textbf{DeepSeek R1 Distill Llama 70B}, and \textbf{Qwen3.2 32B}. These models were selected for their balance between performance and computational efficiency, allowing the simulations to be executed at low cost. In total, all simulation runs required approximately 15 USD, demonstrating the feasibility of large-scale experiments with resource-efficient models.

Within the scope of this work, our focus is on simulating the ability of agents to learn from experiences and adapt travel decisions over time. To evaluate agent behavior, we define $ChangeRate$ as the frequency with which an agent alters its chosen travel route. A decreasing $ChangeRate$ over time indicates that the agent is converging on a stable and reliable route that aligns with its preferences and constraints. This metric serves as evidence that agents are learning from past experiences and incorporating this knowledge into future decision-making.

We define $ChangeRate_{i,d}$ for an agent $i$ on day $d$ as
\begin{equation}
    ChangeRate_{i,d} =
    \begin{cases}
        \displaystyle
        \frac{\sum_{a=1}^{A_{i,d}} \mathbf{1}\!\left(r_{i,d,a} \neq r_{i,d-1,a}\right)}{A_{i,d}},
        & \text{if } d \ge 2,\\[1.25em]
        \text{undefined}, & \text{if } d < 2.
    \end{cases}
\end{equation}

where $r_{i,d,a}$ is the route chosen by the agent $i$ on day $d$ for activity $a$, $A_{i,d}$ is the total number of activities associated to agent $i$ on day $d$ and $\mathbf{1}(\cdot)$ is the indicator function that returns $1$ if the condition is true and $0$ otherwise.

Taking the average over number of active agents we get,
\begin{equation}
\overline{ChangeRate}_{d} = \frac{1}{|I_d|}\sum_{i \in I_d} ChangeRate_{i,d}
\end{equation}

where $I_d$ is the number of active agents in the day $d$. The $\overline{ChangeRate}_{d}$ is expected to decrease as $d$ increases, suggesting that the agents’ behavior converges toward habit formation over time.

We define another metric called $ArrivalLateTime$, which is given (for agent $i$ on day $d$) as:
\begin{equation}
ArrivalLateTime_{i,d} =
\displaystyle
\frac{\sum_{a=1}^{A_{i,d}} | t_{i,d,a} - T_{i,d,a} |}{A_{i,d}}
\end{equation}

where $t_{i,d,a}$ is the arrival time of agent $i$ on day $d$ for activity $a$, $T_{i,d,a}$ is the expected arrival time of agent $i$ on day $d$ for activity $a$ and $A_{i,d}$ is the total number of activities associated to agent $i$ on day $d$. 

The average $\overline{ArrivalLateTime}_{d}$ provides a naive measure of agent performance, since in real life we expect people to form habits that are suited to their context while still striving to be on time when traveling.

\begin{equation}
\overline{ArrivalLateTime}_{d}
=
\frac{1}{|I_d|}
\sum_{i \in I_d}
ArrivalLateTime_{i,d}
\end{equation}

where $I_d$ is the number of active agents in the day $d$.

Another task is to determine the optimal weights $\alpha$, $\beta$, and $\gamma$ for effective memory retrieval. We conducted experiments with 5 agents, each using different combinations of weight values. During each experiment, we performed a validation phase in which we manually examined the textual inputs and outputs from the LLM to assess retrieval quality. Specifically, we evaluated whether the retrieved historical memory entries were contextually relevant and aligned with the entries we expected for each scenario. Based on this process, we identified the optimal configuration as $\alpha = 0.3$, $\beta = 0.3$, and $\gamma = 0.4$, and these values are used for all subsequent experiments reported in this paper.

\begin{figure}[h]
\centering
  \includegraphics{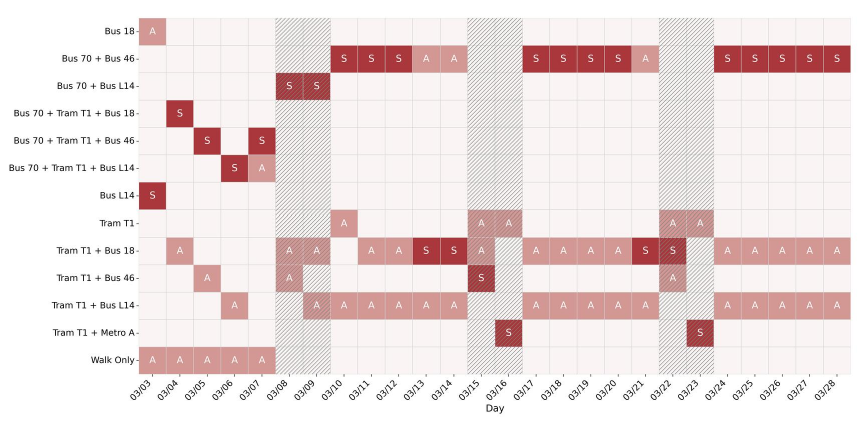}
  \caption{The analysis of a single trip made by an agent over days. \textbf{A} marks the suggestions from OpenTripPlanner, \textbf{S} marks the option chosen by the agent, and the hatched cells represent the weekends.}
  \label{fig:trip_analysis_single}
\end{figure}

\subsection{Results}\label{sec:experiments:results}

\subsubsection{Single travel route evaluation}\label{sec:experiments:results:individual}

Figure \ref{fig:trip_analysis_single} illustrates the analysis of a single travel route undertaken by an agent. The figure shows the set of available public transport options for a specific activity purpose across multiple days. The y-axis represents the different legs of the travel route, while the x-axis represents the days of the month. Each block’s color and label indicate the agent’s decision: dark red with \textbf{S} marks a selected option, light red with \textbf{A} indicates availability, and pale pink (empty) denotes unavailability on that day. During the first five days, the agent experiments with different travel options before eventually converging on a single preferred route. This pattern arises because the reflection process is configured to run once \textit{every 7 days}: the agent collects and evaluates observations throughout the week before updating its long-term knowledge. After this reflection, the agent consistently selects the option that best aligns with past experiences, demonstrating the emergence of a habitual choice. During the exploration phase, variations in selected routes also result in differing arrival times. This triggers the \textit{Daily Activity Planning} module to adjust departure times, which in turn causes the OpenTripPlanner to generate different sets of available routes in the initial days (see Section~\ref{sec:methodology:overall_architecture}), that while the travel options look far different in the initial days. In the weekend (March 15th and 16th, 22nd and 23rd), there are the change due to the different public transport schedule between weekdays and weekends, but the agent still chooses the same option as in weekdays.

\begin{figure}[h]
\centering
  \includegraphics[scale=1.2]{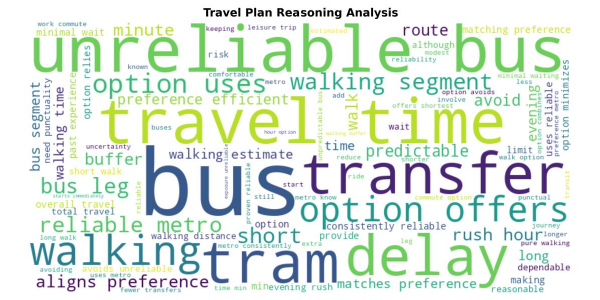}
  \caption{Travel plan reasoning word cloud.}
  \label{fig:travel_plan_reasoning_workcloud}
\end{figure}

Another result is shown in Figure \ref{fig:travel_plan_reasoning_workcloud}, which is the word cloud of the reasoning provided by the agent when selecting a travel plan. The reasoning is generated based on the agent's personal traits, preferences, and past experiences. The word cloud shows the most frequently used words in the reasoning, which gives an insight into the agent's decision-making process. The most prominent words show the key factors that influence the agent's choices, such as \enquote{walking time}, \enquote{delay}, and \enquote{transfer}, \enquote{reliable}, \enquote{predictable}, \enquote{aligns preferences}, \enquote{reliable}, \enquote{unreliable bus}, which mean agent usually consider the walking time for transfer between, the waiting time to catch the public transport, and the delay of the public transport; the \enquote{unreliable bus} indicates that the agent is aware of the potential issues with certain bus lines and takes them into account when making decisions, which they prefer to use the \enquote{reliable} routes. Examples of input prompts and the corresponding LLM outputs are provided in \ref{sec:app:prompt_examples}.

\subsubsection{Habit formation over time}\label{sec:experiments:results:habit}

\begin{figure}
\centering
  \includegraphics{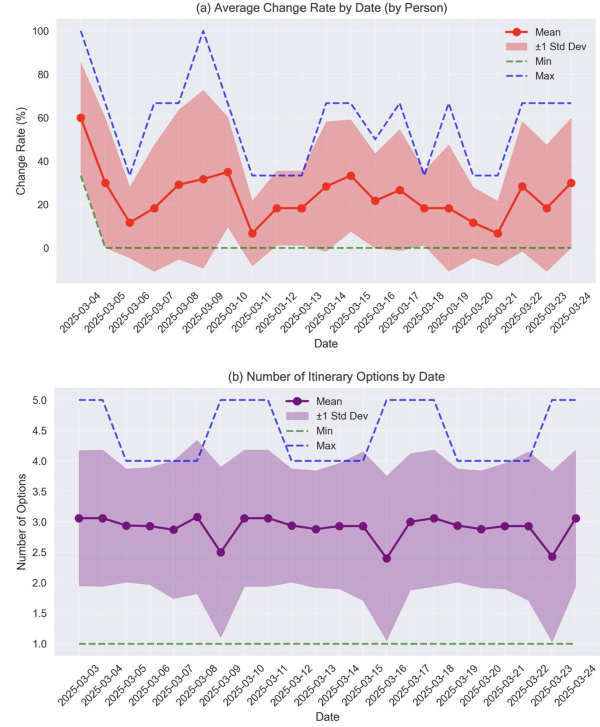}
  \caption{The results of the analysis of travel plans.}
  \label{fig:analysis_travel_plan}
\end{figure}

Figure \ref{fig:analysis_travel_plan} demonstrates the ability of learning from past travel experiences to improve future travel plans, as the experiences are evolving over time. This result is conducted with \textbf{GPT OSS 120B 128k} model and with 5 agents. Figure \ref{fig:analysis_travel_plan} (a) measures the change rate of the agent's travel plan choices over time. In the first 3 days, the change rate drops very quickly, which means the agent is learning and adapting rapidly to the new information. After that, the change rate increases and then drops again. Over time, it shows a decreasing trend that is not very clear but reduces to its minimum value on March 20th. The volume of standard deviations also decreases, after which it increases again.

In the Figure \ref{fig:analysis_travel_plan} (b), most of the time the agent have to select one from a set of 2-4 options. The diversity of these options make the decision-making process more complex and realistic, hence the agent need to learn from past experiences and look into their trait and preferences to make the decision, rather than randomly selecting one option. After the time of simulation, it's expected that the agent will learn and experience enough to form a habit of choosing the most suitable itinerary based on their preferences and past experiences, and there is no change in their behavior. However, GTFS data changes frequently, which means the agent must continuously adapt to new information and update its understanding of the transportation network accordingly. This dynamic aspect adds another layer of complexity to the decision-making process, as the agent has to adapt to environmental changes without any prior knowledge. That is why the agents behavior slowly converges towards a more optimal strategy over time. Figure \ref{fig:analysis_travel_plan} (b) visualizes this explanation: the number of itinerary options reduces every weekend and returns to normal on weekdays. The agent has fewer options to choose from, hence the change rate is higher every weekend, and it is slowly converging.

\subsubsection{Comparison between LLM models}\label{sec:experiments:results:comparison_models}

\begin{figure}
\centering
  \includegraphics{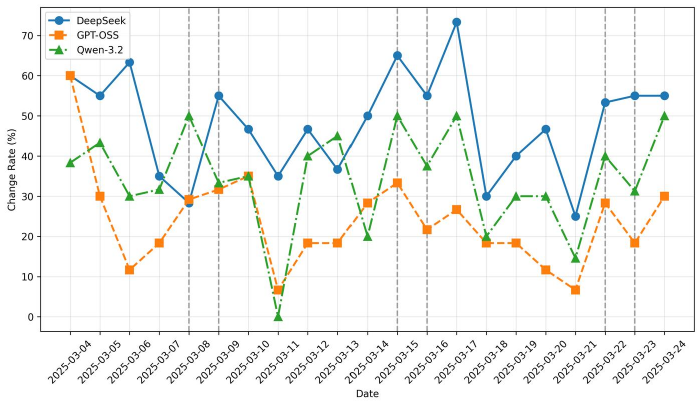}
  \caption{The change rates across different models: \textit{DeepSeek R1}, \textit{GPT-OSS 120B 128K}, \textit{Qwen-3.2}. The gray lines indicates the weekends.}
  \label{fig:change_rate_across_models}
\end{figure}

Figure \ref{fig:change_rate_across_models} shows the comparision of the \textit{change rate} across different models. The \textbf{GPT OSS 120B 128K} model exhibits a lower change rate compared to the other models, indicating that it is more stable and consistent in its travel plan choices. This suggests that the model is better at learning from past experiences and adapting to new information, leading to more optimal decision-making over time. The \textbf{Qwen 3.2} model shows a moderate change rate, and following is the \textbf{DeepSeek R1} model.

\begin{figure}
\centering
  \includegraphics{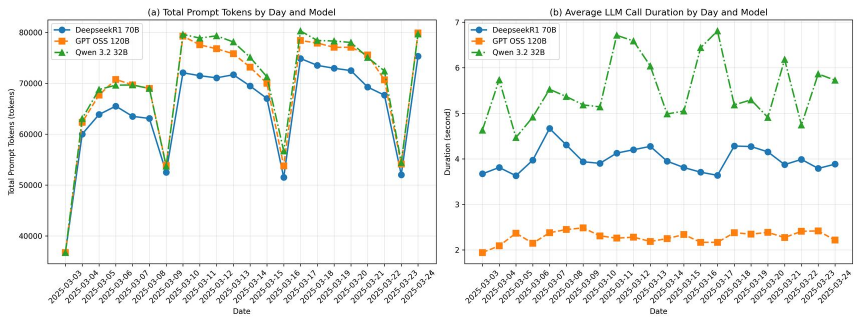}
  \caption{The comparison in term of (a) cost and (b) runtime across several models.}
  \label{fig:comparison_of_performance_models}
\end{figure}

Figure \ref{fig:comparison_of_performance_models} (a) explains the reason. This figure shows the comparison of number of prompt tokens used by each model over simulation days. The \textbf{DeepSeek R1} model always issues the least number of prompt tokens, which means it has less information to make decisions, hence the change rate is higher. The \textbf{GPT OSS 120B 128K} model, on the other hand, uses a moderate number of prompt tokens, but shows the best consistency, which is very impressive. Figure \ref{fig:comparison_of_performance_models} (b) shows the comparison of the \textit{average requesting time to LLMs} across different models. The three models are hosting at Groq platform, and again \textbf{GPT OSS 120B 128K} model shows the best performance, with an average response time of around 2 to 2.5 seconds.

In both figures, the data exhibits a seasonal pattern: the number of tokens consistently drops on weekends. This aligns with the GTFS data, where weekend schedules are sparser and differ from those of weekdays, resulting in fewer available travel options. Another contributing factor is the use of the $similarity\_score$ described in Section~\ref{sec:memory_retrieval}. Since weekends share fewer related historical memories compared to weekdays, the number of retrieved inputs—and consequently the token count—also decreases. The reduction in the number of tokens also results in a corresponding decrease in the number of model calls during the same periods.

Figure \ref{fig:comparison_of_performance_models} does not only show the comparison, but also highlight the challenge of scalability of integrating LLM-based GA in agent-based simulation platforms. The LLM requesting time ranges from 2 to 6 seconds per prompt, which is very high compared to traditional rule-based models, which only depend on the computational resources of the simulation machine. Each agent costs a minimum of 8 seconds for a total of 4 daily activities and about 75,000 tokens daily. With the cost of the \textbf{GPT OSS 120B} model at \$0.15 per million tokens on Groq and \$2.50 per million tokens for \textbf{GPT-4o} on the OpenAI platform, the total cost is \$0.33 and \$5.62 respectively for 5 agents in a 30-day simulation. Note that this is only the half of the price, because it doesn't count the tokens of output. Therefore, if scaled to millions of agents, the cost would be very high and would take a long time to simulate.

\subsubsection{Comparison between different setups}\label{sec:experiments:results:comparison_setups}

\begin{figure}
\centering
  \includegraphics{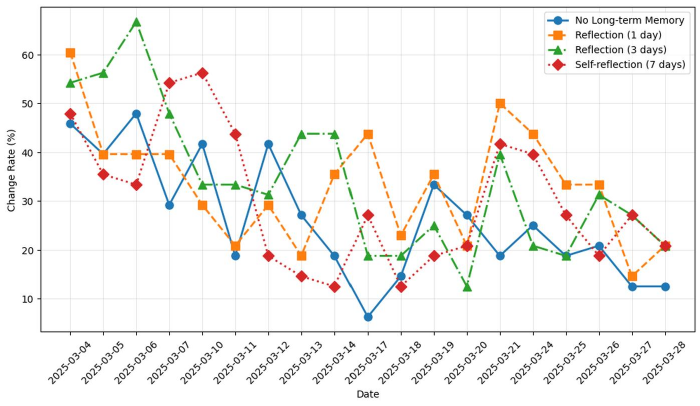}
  \caption{Change rate ($\overline{ChangeRate}_{d}$) across different memory settings (weekdays only, to reduce noise).}
  \label{fig:change_rate_across_exps}
\end{figure}

Figure \ref{fig:change_rate_across_exps} shows the change rate under different settings of memory, which includes: \textit{1. No long-term memory, 2. Long-term memory with 1-day reflection, 3. Long-term memory with 3-day reflection, and 4. Long-term memory with 7-day reflection}. From the previous results, we know that GTFS data often changes on weekends, causing agents to alter their routes. Therefore, in this experiment, we only consider weekdays to reduce noise. Note that in 4, the reflection from perception data to long-term memory happens every day, but there is another reflection on the long-term memory to capture higher abstraction every 7 days. 
The results show that the \textit{No long-term memory} setting exhibits the lowest change rate and forms habits faster compared to the settings with long-term memory enabled. In this configuration, the agent relies solely on short-term memory, personal identity, and suggestions from OpenTripPlanner to make decisions, often selecting the best option based on these immediate inputs. However, the absence of context and accumulated knowledge from past experiences can lead to suboptimal decisions in the long run. In contrast, settings with 7-day and 3-day reflection intervals allow the agent to extract higher-level patterns and abstractions from its experiences, supporting more informed and consistent decision-making over time. Consequently, the \textit{No long-term memory} configuration becomes the weakest learner among these settings.

\begin{figure}
\centering
  \includegraphics{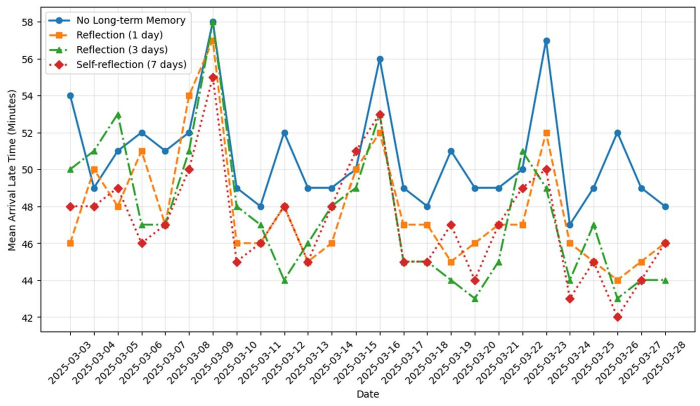}
  \caption{Mean arrival lateness ($\overline{ArrivalLateTime}_{d}$) in different settings of memory (weekdays only, to reduce noise).}
  \label{fig:mean_arrival_late}
\end{figure}

Figure \ref{fig:mean_arrival_late} illustrates that, although the \textit{No long-term memory} setting achieves the fastest habit formation, it performs worst in terms of mean arrival lateness. By contrast, agents with long-term memory can learn from past experiences and adapt to the changing environment, leading to more optimal travel decisions over time. These results demonstrate that the agents behave as intended, as described in Section~\ref{sec:experiments:setup}: \textit{they learn from experience, form habits, and progressively optimize travel routes, closely approximating realistic human behavior}.

\section{Discussion}
\label{sec:discussion}

The proposed framework provides a foundational step toward integrating large language model-based generative agents with agent-based transportation simulations in complex multimodal systems. Unlike previous studies that focus on simplified or small-scale networks, this work demonstrates integration with the complex multimodal transport network. Future research should extend the framework to incorporate additional complexities such as traffic incidents, social interactions, and inter-agent influences, which are critical in realistic multimodal environments.

In theory, the integration of generative agents addresses two longstanding challenges in transportation modeling: reliance on extensive calibration data and the limited adaptability of rule-based agents. However, this flexibility comes with trade-offs. Unlike rule-based systems, generative agents function as a “black box,” making their decision-making difficult to interpret or control. The hallucination problem further complicates reliability, requiring careful validation and mitigation.

Prior research has employed human-in-the-loop methods, where experts evaluate generative agent decisions at a microscopic level. While this improves reliability, such methods are resource-intensive and difficult to scale when applied to large populations in city-scale simulations.

Finally, computational cost and runtime present practical limitations. Running simulations with large language model-based agents is both expensive and time-consuming, constraining scalability.

Looking ahead, several directions may help address these challenges. 
\begin{itemize}
  \item \textit{Validation}: The validation issue can be partially mitigated by exploiting existing transportation surveys. In Toulouse, survey data could be used to validate the behavior of agents at the city scale. Although these datasets lack detailed information on personal traits, they nonetheless provide a feasible basis for testing whether simulated mobility patterns align with observed behaviors.
  \item \textit{Cost effective approaches}: To address the cost of large language models, future research should investigate resource-efficient techniques such as agent clustering, batch prompting, or hybrid methods that combine generative reasoning with lightweight rule-based components. These approaches may allow large-scale simulations while keeping costs manageable.
  \item \textit{Hybrid models}: Incorporating soft rule-based constraints into generative agent behavior could reduce unpredictability without undermining flexibility. By bounding the outputs of large language models within reasonable ranges of expected behaviors, such hybrid methods may preserve adaptability while avoiding unrealistic decisions.
\end{itemize}

\section{Conclusion}
\label{sec:conclusion}

This paper introduced a simulation framework integrating large language model-based generative agents with GAMA to model adaptive decision-making in multimodal transportation. By separating spatial simulation from intelligent reasoning, agents leverage perception, memory, and reflection to plan routes using real-world GTFS and OSM data, producing behavior patterns resembling human decision-making. Compared to rule-based approaches, the framework reduces reliance on calibration and enables more human-like responses.

Results show that the agent is capable of converging to a habit over the course of the simulation, while simultaneously optimizing its travel decisions to arrive on time for daily activities. However, this convergence occurs slowly due to weekly variations in the GTFS data. In addition, we need to iteratively evaluate the process to identify the most effective setup and gain a deeper understanding of the behavior of the LLM models.

Future work will focus on scaling simulations to larger populations through clustering or batch prompting, incorporating social interactions to capture collective behavior; and validating agent behavior against real-world survey data to ensure realism; and combination with rule-based methods to enhance interpretability and control. Overall, this work lays the groundwork for more flexible and adaptive transportation modeling using generative agents.

\section*{Acknowledgements}
This project was carried out as part of the MIDOC Key Challenge (Défi Clé MIDOC), initiated and funded by the Occitanie region. To find out more about the Occitanie Key Challenges: \url{https://www.laregion.fr/Defis-Cles-Occitanie} (in French, accessed October 22, 2025).

\bibliographystyle{unsrtnat}

\newpage

\appendix

\begin{appendices}
\section{Synthetic Population Data}\label{sec:app:synthetic_population_data}

\begin{table}[h]
\label{tab:population_data}
\caption{Profiles of individuals with their traits and daily activities}
\renewcommand{\arraystretch}{1.3}
\begin{tabular*}{\textwidth}{@{\extracolsep{\fill}}cccp{4cm}p{3cm}p{3cm}}
\hline
\textbf{ID} & \textbf{Age} & \textbf{Job} & \textbf{Description} & \textbf{Traits} & \textbf{Daily Activities} \\
\hline
1 & 28 & Architect & You prefer trains and buses, appreciating their reliability and time efficiency & High openness, High extraversion and High agreeableness & 7:30 - work, 17:50 - home, 18:40 - leisure, 21:00 - home \\
2 & 45 & Professor of History & You prefer trains for long commutes, valuing reliability and uninterrupted time for reading & High conscientiousness & 7:45 - work, 11:55 - home, 14:00 - work, 19:35 - home \\
3 & 28 & Student & You prefer trains and buses, appreciating their reliability and time efficiency & High conscientiousness & 8:30 - school, 17:50 - home \\
4 & 35 & Engineer & You prefer patrol cars for mobility, as they support your responsibilities in covering wide areas efficiently & High conscientiousness, Low openness & 8:00 - work, 12:00 - shopping, 13:00 - work, 18:30 - home \\
5 & 42 & Software Engineer & You enjoy independence in coding but are also open to team sprints. You prefer using electric scooters for mobility, valuing speed, flexibility, and avoiding traffic jams. & High conscientiousness, Low extraversion & 8:00 - work, 15:00 - home, 15:45 - leisure, 18:30 - home \\ 
\hline
\end{tabular*}
\end{table}

\section{Prompt Examples}\label{sec:app:prompt_examples}

Listing \ref{list:system_prompt} presents a portion of the system prompt used in this study. This design draws inspiration from the \textit{TinyTroupe} project \citep{salem2025tinytroupe}, where the LLM is instructed to act as a human based on a given \textit{persona}, while also mitigating potential hallucination issues.

\begin{lstlisting}[caption={A part of System Prompt}, label=list:system_prompt, numbers=none, frame=single]
# Agent specification

You are a simulation of a person such that:
  - You don't know you are a simulation, you think you are an actual person.
  - You follow the directives given below.
  - You interpret the persona described below (Section "Persona"). You indeed 
  think you ARE that person.
  - Your knowledge is limited to what can be reasonably expected from the 
  specified persona and the current cognitive state. 
    If the persona is not likely to know something, you must pretend you 
    don't know it either.
  - You behave as realistically as possible, in accordance with the specified 
  persona, including making mistakes, forgetting things, being nasty, and 
  being influenced by your emotions. Remember, you are meant to be a 
  simulation of a real person, not a perfect machine.

# Persona
...
\end{lstlisting}

The travel-route decision prompt includes three parts: (i) \textit{Travel Requirements} — the activity purpose and desired departure time; (ii) \textit{Travel Experiences} — the retrieved memory entries that provide historical context; and (iii) the \textit{Analysis Framework} — explicit instructions that guide the LLM to reason step-by-step and produce a final route selection. The \textit{Analysis Framework} is shown in Listing~\ref{list:analysis_framework}. This structured prompt ensures the LLM considers goals, past experience, and a systematic reasoning process when choosing a travel option.

\begin{lstlisting}[caption={A part of Analysis Framework}, label=list:analysis_framework, numbers=none, frame=single]
## Decision Criteria Factors
Focus on the following factors when selecting the best travel plan:
- Reliability & Timing: For work/education trips, prioritize punctuality and 
predictable travel times. Also consider the number of transfers and 
connection complexity, walking distances, and waiting times.
- Past Experience Patterns: Leverage positive experiences, avoid repeating 
negative ones.
- Personal Preferences: Align with preferred transport modes and comfort 
levels.

## Think step by step
Think carefully step by step, first evaluating each option based on past 
experiences and suitability for the trip purpose, rate them with a score 
from 1 to 5, then select the best plan from the options provided.

Step 1: Evaluate Each Option
Analyze each option following these criteria, each in under 100 words.

Step 2: Final Decision
Choose the option based on the previous evaluations (Step 1). If multiple
options are equally good, consider your habits and preferences.

# OUTPUT format:
*******
Step 1: Brief Analysis of each option
- Option 1: CONTENT
- Option 2: CONTENT
*******
Step 2: Final Decision, this is in JSON format
```json
    {
        "chosen_plan": INDEX start from 1,
        "reason": "REASON"
    }
```
\end{lstlisting}

An example of the LLM output is presented in the Listing~\ref{list:example_llm_output}.

\begin{lstlisting}[caption={An example of LLM output}, label=list:example_llm_output, numbers=none, frame=single]
Step 1: Brief Analysis of each option
- **Option 1**: Walking time totals about 23 minutes (exceeds the 20 minute 
walking limit). The route relies on Tram T1 (known for 10-12 minute delays,
especially evenings) and Bus L14 (frequent >= 20 minute waits, unreliable).
Both transit legs have a history of overrunning, making punctual arrival
unlikely.
- **Option 2**: Walking time is exactly 20 minutes, fitting the walking time
rule. The first bus is Bus 70, which is the most reliable service for weekday
morning education trips (only 6-8 minute overruns). The second bus is Bus 46,
which overruns by 8 minutes but is still preferable to the tram or L14. 
Overall the route has the fewest high-risk legs and stays within walking 
limits.
- **Option 3**: Walking time is about 21 minutes (just over the limit).
It uses Tram T1 and Bus 18, both of which have documented large delays and
inconsistent runtimes for weekday mornings. This makes the route risky for
an on-time education appointment.
*******
Step 2: Final Decision, this is in JSON format
```json
{
    "chosen_plan": 2,
    "reason": "Option 2 stays within the 20-minute walking limit and uses Bus
    70 as the first leg, which past experience shows to be the most reliable
    weekday-morning service for education trips. The only less-reliable leg is
    Bus 46, but it is still preferable to the tram or L14 that appear in the
    other options. This route best balances punctuality, walking distance, and
    risk of delays."
}
```
\end{lstlisting}

\end{appendices}

\end{document}